\documentclass[12pt,showpacs,floatfix,preprintnumbers,nofootinbib]{iopart}
\usepackage{amsopn}
\usepackage{amssymb}
\usepackage{hyperref}
\usepackage{color}
\usepackage{graphicx}
\usepackage{acronym,xspace}
\usepackage{url}

\newcommand{\be}{\begin{equation}}
\newcommand{\ee}{\end{equation}}
\newcommand{\bel}[1]{\begin{equation}\label{#1}}
\newcommand{\ba}{\begin{eqnarray}}
\newcommand{\ea}{\end{eqnarray}}
\newcommand{\bal}[1]{\begin{eqnarray}\label{#1}}

\newcommand{\vtheta}{\vec{\theta}}
\newcommand{\inner}[2]{{\left\langle #1 \middle| #2 \right\rangle}}
\newcommand{\order}[1]{{\mathcal{O}\left( #1 \right)}}
\newcommand{\dd}{\ensuremath{\mathrm{d}}}
\newcommand{\sub}[1]{\ensuremath{_\mathrm{#1}}}
\newcommand{\super}[1]{\ensuremath{^\mathrm{#1}}}

\DeclareMathOperator{\Exn}{E}

\renewcommand{\Re}{\operatorname{Re}}

\begin{document}
\title[Parameter estimation with abruptly terminated waveforms]
{Parameter estimation on compact binary coalescences with abruptly terminating gravitational waveforms}

\date{\today}

\author{Ilya Mandel$^{1,2}$ \footnote{email:imandel@star.sr.bham.ac.uk}, Christopher P L Berry$^1$, Frank Ohme$^3$, Stephen Fairhurst$^3$, Will M Farr$^1$}

\address{$^1$ School of Physics and Astronomy, University of Birmingham, Edgbaston, Birmingham B15 2TT, United Kingdom}
\address{$^2$ National Institute for Theoretical Physics (NITheP), Western Cape, South Africa}
\address{$^3$ School of Physics and Astronomy, Cardiff University, Cardiff, CF24 3AA, United Kingdom}


\begin{abstract}
Gravitational-wave astronomy seeks to extract information about astrophysical systems from the gravitational-wave signals they emit. For coalescing compact-binary sources this requires accurate model templates for the inspiral and, potentially, the subsequent merger and ringdown. Models with frequency-domain waveforms that terminate abruptly in the sensitive band of the detector are often used for parameter-estimation studies. We show that the abrupt waveform termination contains significant information that affects parameter-estimation accuracy. If the sharp cutoff is not physically motivated, this extra information can lead to misleadingly good accuracy claims. We also show that using waveforms with a cutoff as templates to recover complete signals can lead to biases in parameter estimates. We evaluate when the information content in the cutoff is likely to be important in both cases. We also point out that the standard Fisher matrix formalism, frequently employed for approximately predicting parameter-estimation accuracy, cannot properly incorporate an abrupt cutoff that is present in both signals and templates; this observation explains some previously unexpected results found in the literature. These effects emphasize the importance of using complete waveforms with accurate merger and ringdown phases for parameter estimation.
\end{abstract}


\maketitle

\acrodef{GW}{Gravitational-wave}
\acrodef{BH}{black hole}
\acrodef{NS}{neutron star}
\acrodef{PN}{post-Newtonian}
\acrodef{SNR}{signal-to-noise ratio}
\acrodef{ISCO}{innermost stable circular orbit}
\acrodef{FIM}{Fisher information matrix}

\newcommand{\PN}[0]{\ac{PN}\xspace}
\newcommand{\BH}[0]{\ac{BH}\xspace}
\newcommand{\NS}[0]{\ac{NS}\xspace}
\newcommand{\GW}[0]{\ac{GW}\xspace}
\newcommand{\SNR}[0]{\ac{SNR}\xspace}
\newcommand{\ISCO}[0]{\ac{ISCO}\xspace}
\newcommand{\FIM}[0]{\ac{FIM}\xspace}

\section{Introduction}

\GW astronomy endeavours to infer the properties of astrophysical systems from the gravitational radiation they emit. For ground-based detectors, such as the Laser Interferometer Gravitational-wave Observatory (LIGO) and Virgo \cite{LIGO,Virgo}, a principal \GW source are binaries consisting of neutron stars or stellar-mass black holes that inspiral and eventually coalesce as GWs carry away energy and angular momentum. Parameters of interest for these systems include the component masses and spins and the location and orientation of the binary. With the upcoming advanced generation of \GW detectors \cite{AdvLIGO,AdvVirgo}, which are expected to make the first detections of coalescing black-hole and neutron-star binaries \cite{scenarios,ratesdoc}, efforts to predict parameter-estimation accuracy have intensified.

Over the past two decades, a variety of techniques have been used for predicting the accuracy with which parameters can be extracted from a detected \GW signal. The \FIM formalism has been particularly popular because of its low computational cost and ease of use \cite{Vallisneri:2008}. Dozens of studies have used the \FIM tool, including the classic work of \cite{Finn:1992wt,CutlerFlanagan:1994,PoissonWill:1995}, ranging in applications from tests of GR \cite{Rodriguez:2012,PaiArun:2013} to sky-localization predictions for multi-messenger astronomy \cite{Fairhurst:2009, Fairhurst:2011, Grover:2013}. More recently, alternatives to the \FIM formalism have been considered, e.g., \cite{Hannam:2013,Ohme:2013,Cho:2013}. Finally, as computational resources have expanded, more costly Bayesian techniques \cite{Jaynes} have been employed to compute the full posterior probability density functions of the signal parameters. These techniques, based on stochastically sampling the parameter space with methods such as Markov chain Monte Carlo and nested sampling, have been used to consider measurements of masses and spins for different classes of systems and the sky-localization ability of different network configurations, e.g., \cite{vanderSluys:2008a,VeitchVecchio:2009,Raymond:2010,Veitch:2012,Rodriguez:2013BNS}. 

Despite the differences in methodology, all of the studies referenced above, and
many others, share one common feature: they use waveforms that terminate
abruptly in the band of the detectors. We do not expect most real \GW signals
to exhibit such a steep falloff, but instead that they evolve smoothly through inspiral,
merger and ringdown phases. However, accurate waveforms that included all
phases of the \GW signal were not available until recent advances in numerical
relativity (see \cite{Sperhake:2011,Pfeiffer:2012} for recent reviews) allowed analytical waveform families to be constructed by
calibrating against numerical results, e.g.,
\cite{Ohme:2012,Santamaria:2010,Ajith:2011b,Pan:2011,Taracchini:2012,
Taracchini:2013,Damour:2013}. Meanwhile, inspiral-only waveforms based on the
post-Newtonian expansion and terminating at the \ISCO have been known for many
years \cite{Blanchet:2014} and are computationally inexpensive to calculate. Consequently,
it was natural for the early studies to make use of these waveforms. Even now,
there are cases where it is beneficial to use post-Newtonian waveforms with an
abrupt termination.

Frequency-domain waveforms based on the stationary-phase approximation
\cite{Damour2001,Damour2002,PNwaveforms:2009} are particularly well suited to
both analytical and numerical studies. Such a waveform, terminated at the \ISCO, can be written
as
\ba
\tilde{h}(f) &=& A(f) \exp\left[i \Psi(f)\right] H(f\sub{ISCO}-f) \label{eq:waveformmodel} \\
 &\equiv& \tilde{h}^0(f) H(f\sub{ISCO}-f),
\label{eq:cutoff}
\ea
where $f\sub{ISCO}$ is the GW frequency at the \ISCO, and $H$ is the Heaviside step function. These have been generally used for both \FIM calculations and parameter-estimation studies (as discussed above), with a few notable exceptions including \cite{VitaleZanolin:2010,S6PE,NINJA2}, as well as for gravitational-wave searches \cite{S6lowmass, findchirppaper, ihope}. However, the impact of the step function, i.e., the sharp waveform cutoff, is typically ignored in these applications.
 
In this paper, we investigate in detail the effect of using waveforms with a sharp cutoff in the frequency domain on parameter recovery. After briefly recalling the likelihood and \FIM formalism (Sec.~\ref{sec:FIM}), we begin by considering the case where both the true signal and the waveforms used to recover it terminate abruptly. We show that the abrupt termination significantly alters the information content of the signal. In particular, while the accuracy of measurement typically scales inversely with the \SNR, parameters associated with an abrupt cutoff can be measured with an uncertainty proportional to the square of the inverse \SNR (Sec.~\ref{sec:SNR-scaling}). We describe the regime in which the abrupt waveform termination significantly impacts parameter-estimation accuracy (Sec.~\ref{sec:abrupt-sig}) and derive the likelihood function for a data set given a model with a sharp waveform cutoff (Sec.~\ref{sec:analytic-approx}). Subsequently, we consider the impact of the abrupt waveform termination on the \FIM formalism, and explain the apparent violation of the Cram{\'e}r--Rao bound found by \cite{Rodriguez:2013}, whose Bayesian confidence intervals on mass parameters were a few times smaller than those predicted by their \FIM (Sec.~\ref{sec:new_FIM}).

Then, in Sec.~\ref{sec:fullsignal}, we investigate the impact of using template waveforms with an abrupt cutoff in searching for, and estimating the parameters of, signals which extend smoothly through merger and ringdown. We show that, at leading order, the parameter accuracies given by the \FIM are correct if inspiral-only information is used, although the presence of a merger and ringdown can lead to a systematic offset in the recovered parameters and the use of full inspiral--merger--ringdown waveforms for the analysis could allow for more accurate parameter estimation.
We evaluate this bias in recovered parameters and identify the regime where cutoff waveforms introduce significant bias into signal recovery.

While we limit our discussion to the specific application to \GW signals, we hope it is of interest to other fields which employ similar parameter-estimation techniques.

\section{Likelihood and Fisher information matrix}
\label{sec:FIM}

In \GW astronomy, the observed data $d$ is generally modeled as a sum of a waveform $h$, which is a function of system parameters $\vtheta_0$ according to an assumed waveform model, and an additive stationary and Gaussian noise $n$:
\be
d(t)= h(\vtheta_0; t) + n(t), \label{eq:data}
\ee
with the frequency-domain expectation value of $n$ being
\be
\Exn [\tilde{n}(f') \tilde{n}^*(f)] = \delta(f-f') S_n(f), \label{eq:Sn_def} 
\ee
where $\tilde{x}(f)$ is the Fourier transform of $x(t)$, $\tilde{x}^*(f)$ is
the complex conjugate of $\tilde{x}(f)$, and $S_n(f)$ is the frequency-dependent
noise power spectral density.

Both the detection of a \GW signal and the extraction of astrophysical source parameters
rely on calculating 
the likelihood $L$ of observing a data set $d$ given a waveform model $h(\vtheta)$. The logarithm of this likelihood, ignoring an additive constant, is given by
\be
\log L (\vtheta) \equiv \log p (d|\vtheta) = - \frac{1}{2} \inner{d-h(\vtheta)}{d-h(\vtheta)}, \label{eq:logL_def}
\ee
where the inner product is defined as
\be
\inner{a}{b} = 4 \,\Re \int_0^\infty \frac{\tilde{a}(f) \tilde{b}^*(f)}{S_n(f)} \,\dd f.
\ee

The expectation value of the likelihood over noise realizations $\Exn [\log L(\vtheta)]$ can be expanded around its value at the signal parameters in a Taylor series {\it if} the likelihood is a smooth, differentiable function at the signal parameters; as we will show below, this condition fails for abruptly terminated templates. The lowest non-vanishing contribution to this Taylor expansion comes at the quadratic order in parameter deviations $\Delta \vtheta \equiv \vtheta - \vtheta_0$, and is proportional to the \FIM $\Gamma_{ij}$ which is defined as
\be
\Gamma_{ij} (\vtheta_0) = \Exn \left[ -\frac{\partial}{\partial \theta_i} \frac{\partial}{\partial \theta_j} \log L (\vtheta)\right]. 
\ee
Here, the expectation value is taken over the possible measurements given true parameters $\vtheta_0$, i.e., over possible noise realizations $n$. In the \GW data-analysis context, the \FIM is given by
\ba\label{Gamma}
\Gamma_{ij} (\vec{\theta_0}) &=& \Exn [-\inner{d}{h_{,ij}}+\inner{h_{,i}}{h_{,j}}+\inner{h}{h_{,ij}}] \nonumber \\ 
 &=& \inner{h_{,i}}{h_{,j}},
\ea
where $h_{,i} \equiv \partial h / \partial \theta_i$, and we have used the fact that the expectation value of the noise vanishes.

In the high-\SNR limit, parameters are constrained sufficiently well that only small variations compared to their true values are probable, hence the linear-signal approximation should hold, $h(\vtheta) \approx h(\vtheta_0) + \sum_i  h_{,i}  \Delta \theta_i$. In the linear-signal, high-\SNR approximation, the covariance matrix $\Sigma_{ij}$ can be approximated as the inverse \FIM \cite{Vallisneri:2008}:
\be
\Sigma_{ij} \simeq \left( \Gamma^{-1} \right)_{ij} = \inner{h_{,i}}{h_{,j}}^{-1}.
\ee
What constitutes high \SNR depends upon the signal, hence this approximation must be checked for each source type (cf.\ \cite{Berry:2013}). 

Regardless of the validity of this approximation, the \FIM always yields the Cram{\'e}r--Rao lower bound (see, e.g., \cite{Vallisneri:2008}) on the expectation over noise realizations of the variance of any unbiased estimator for $\vtheta_0$ at fixed $\vtheta_0$:
\be
\left | \Sigma_{ij} \right | \geq \left| \left( \Gamma^{-1} \right)_{ij} \right|.
\ee
Comparisons of \FIM results against Monte Carlo studies \cite{Cokelaer:2008} and Bayesian parameter estimation methods \cite{Rodriguez:2013} have shown that the priors used in Bayesian analysis provide additional information that, if not included in \FIM analysis, can lead to apparent violations of the Cram{\'e}r--Rao bound.

With the introduction of a sharp cutoff in the waveform, the picture changes dramatically. Rodriguez {\textit{et al}}.\ \cite{Rodriguez:2013} found that even when priors are accounted for, Bayesian analysis recovers confidence intervals on mass parameters that are a few times smaller than those predicted by the \FIM, in apparent violation of the Cram{\'e}r--Rao bound. However, for binary merger waveforms, the \ISCO frequency depends on the parameters of the waveform, in particular the total mass. Thus, one should take into account the changing cutoff frequency when calculating the \FIM. This is generally not done, and holds the key to understanding the apparent violation of the Cram{\'e}r--Rao bound: there is extra information present in the sharp waveform cutoff, which is available to Bayesian techniques but not incorporated in these \FIM calculations. We show how to add this information to the \FIM in Sec.~\ref{sec:new_FIM}.

\section{Effect of a sharp frequency cutoff} \label{sec:cutoff}

\subsection{Accuracy of measuring a sharp cutoff}\label{sec:SNR-scaling}

Although the accuracy of parameter estimation typically scales inversely with the \SNR in the high-\SNR limit, this does not happen when the waveform ends abruptly in-band. To demonstrate this, we first consider a toy scenario. We assume a waveform family of the form given in (\ref{eq:waveformmodel}), where all parameters are fixed except for the cutoff frequency. This simplifies the inner product of (\ref{eq:logL_def}) tremendously, because the only nonvanishing contribution comes from the frequency interval in which one waveform has terminated already but the other one has not (cf.~\cite{Ohme:2013}).


Let the cutoff frequency of the measured signal be $f_0$ and the template be terminated at $f\sub{cut}$. 
We refrain from using the symbol $f\sub{ISCO}$ for now, as we first neglect correlations with other physical parameters and view $f\sub{cut}$ as the only free parameter. 
Let us further assume that the noise realization happens to be exactly zero for the particular measurement we undertook; this is done only to simplify calculations, as measurement accuracy is not affected by the choice of noise realization in the high-\SNR limit. Equivalently, the same results could be obtained by averaging over an ensemble of noise realisations. The first-order expansion of the parameter-dependent part of (\ref{eq:logL_def}) then reads
\ba
-2 \log L(\vtheta) =  \inner{\Delta h}{\Delta h} &=& 4 \left \vert \int_{f_0}^{f\sub{cut}} \zeta(f) \,\dd f\right \vert \nonumber \\
  &\approx& 4 \zeta(f_0) \left \vert  f_0 - f\sub{cut} \right \vert, \label{eq:logL_noiseless}
\ea
where $\Delta h \equiv h(\vtheta_0) - h(\vtheta)$ is the waveform difference, in this example $\Delta h = h^0(f) H(f_0-f) - h^0(f) H(f\sub{cut}-f)$, and the noise weighted signal power $\zeta(f)$ defined as
\be
\label{eq:zeta}
\zeta(f) = \frac{|A(f)|^2}{S_n(f)}.
\ee
The log likelihood scales {\it linearly}  with the {\it absolute value} of $ f_0 - f\sub{cut}$; this is different from the usual quadratic scaling of the log likelihood with parameter variations, which permits a Taylor expansion and leads to the \FIM derivation.

As $\zeta$ depends quadratically on the overall signal amplitude, we can write
\be
4 \zeta(f_0) \equiv \kappa(f_0) \rho^2,
\ee
where $\rho = \sqrt{\inner{h}{h×}}$ is the \SNR and $\kappa$ is some ($f_0$
dependent) constant.\footnote{As an example, consider the (unrealistically) simple scenario where $\zeta(f)$ is constant for $0 \leq f \leq f_0$; in this case $\rho^2 = 4\zeta f_0$, hence $\kappa = 1/f_0$.}

With these simplifications, we can infer the posterior probability of $f\sub{cut}$ being the correct parameter, assuming a uniform prior on $f\sub{cut} \in [0, \infty]$.
The resulting probability density reads
\be
p(f\sub{cut}|d) = \frac{1}{Z} \exp\left(- \frac{\kappa \rho^2 |f\sub{cut}-f_0|}{2}\right),
\ee
where the normalization is given by
\be
 Z = \frac{2}{\kappa \rho^2} \left[ 2 - \exp\left(- \frac{\kappa \rho^2 f_0}{2} \right)\right].
\ee
Calculating the expectation values over this distribution yields
\ba
 \Exn [f\sub{cut}] 
 &=& f_0 + \order{\exp\left(-\frac{\kappa \rho^2 f_0}{2} \right)}
\ea
and
\ba \Exn \left[f\sub{cut}^2\right] 
 &=& f_0^2 + \frac{8}{\kappa^2 \rho^4} + \order{\exp\left(- \frac{\kappa \rho^2 f_0}{2} \right)}.
\ea
The variance is 
\ba
\sigma_{f\sub{cut}}^2 &=& \Exn \left[f\sub{cut}^2\right] - \left(\Exn [f\sub{cut}]\right)^2 \nonumber \\ 
&=& \frac{8}{\kappa^2 \rho^4} + \order{\exp\left(-\frac{\kappa \rho^2 f_0}{2} \right)}; \label{eq:variance_fcut}
\ea
thus, we find that the uncertainty in the measurement of $f\sub{cut}$ actually scales inversely with the square of the \SNR.

This scaling of the measurement uncertainty with the inverse of the square of \SNR, rather than the inverse of the \SNR is, in fact, typical for problems with a sharp cutoff. For example, consider the problem of finding the minimum or maximum of some distribution (say, neutron-star spins) given $N$ observations. Although generally the uncertainty in distribution parameters (e.g., the distribution mean) scales as $1/\sqrt{N}$, the accuracy with which a sharp cutoff can be measured scales as $1/N$ (cf.\ \cite{Chakrabarty:2003}).

\subsection{Significance of an abrupt cutoff} \label{sec:abrupt-sig}

In real \GW searches, the cutoff frequency is often conveniently taken to be the frequency of the innermost stable circular orbit of a test particle orbiting a non-spinning black hole,
\be
 f\sub{cut} \equiv f\sub{ISCO}= \frac{1}{6^{3/2}\pi M}, \label{eq:fISCO}
 \ee
where $M = m_1 + m_2$ is identified as the total mass of the system. For black-hole binaries with two $10 M_\odot$ components, $f\sub{ISCO} = 220~\mathrm{Hz}$ is in the band of initial and advanced detectors; for binary neutron-star systems, the \ISCO frequency of $\sim 1500~\mathrm{Hz}$ is sufficiently high that to be effectively out of band. 
 
The accuracy of measuring $f\sub{ISCO}$ depends on where this frequency falls on the detector's noise spectrum. However, it is usually not included in the set of independent parameters; rather, $f\sub{ISCO}$ is defined by the correlation with mass-dependent parameters. We can nevertheless use the derivation of the previous section to gauge where such a parameter-dependent cutoff frequency is relevant.
 
In the previous section, we calculated the measurement uncertainty of
$f\sub{ISCO}$, given by (\ref{eq:variance_fcut}), under the assumption that all
other parameters are perfectly known.  We denote this uncertainty
$\sigma_{f\sub{ISCO}}\super{abrupt}$ to highlight the inclusion of the cutoff,
and re-express it as
\begin{equation}
\sigma_{f\sub{ISCO}}\super{abrupt} \simeq \frac{1}{\sqrt{2} \zeta(f\sub{ISCO})}.
\end{equation}
Assuming perfect knowledge of the other parameters causes this to be an underestimation of the true uncertainty in determining the \ISCO frequency.

Alternatively, we can compute the predicted parameter-estimation accuracy without taking the abrupt cutoff into account by using the naive, inspiral-only \FIM $\Gamma_{ij}$ that ignores the presence of the cutoff (see Sec.~\ref{sec:new_FIM}) and convert this to an estimate of the accuracy with which $f\sub{ISCO}$ could be measured using the other parameters. We denote this $\sigma_{f\sub{ISCO}}\super{naive}$. If the total mass $M$ and the mass ratio $q=m_2/m_1$ are used to parametrize the binary with non-spinning components, the only non-vanishing partial derivative of $f\sub{ISCO}$ with respect to the parameters is
\be
\frac{\partial f\sub{ISCO}}{\partial M} = - \frac{f\sub{ISCO}}{M}.
\ee
Consequently,
\be
\sigma\super{naive}_{f\sub{ISCO}} = \left| \frac{\partial f\sub{ISCO}}{\partial M} \right| \sigma_M =  {f\sub{ISCO}} \frac{\sigma_{M}}{M}, 
\ee
where $\sigma_{M}$ is the naive \FIM prediction for the mass measurement uncertainty, $\sigma_M^2 = (\Gamma^{-1})_{MM}$.

When the condition
\begin{equation}
\sigma\super{naive}_{f\sub{ISCO}} \ll \sigma\super{abrupt}_{f\sub{ISCO}}
\label{eq:ignore_cutoff}
\end{equation}
holds, there is little information content in the waveform cutoff, and it is generally safe to ignore the impact of the cutoff on parameter estimation. However, if this condition is violated, the information from the abrupt cutoff can significantly reduce parameter-estimation uncertainty. For example, for a noise spectrum roughly representative of initial LIGO sensitivity \cite{Damour2001}, which was
used in \cite{Rodriguez:2013}, $\sigma\super{naive}_{f\sub{ISCO}} \approx \sigma\super{abrupt}_{f\sub{ISCO}}$ at a total mass of approximately $15 M_\odot$ at $\rho = 10$ (higher \SNR makes the abrupt cutoff significant at lower masses). Thus, for this noise spectrum, (\ref{eq:ignore_cutoff}) holds for low-mass binaries such as neutron-star--neutron-star and low mass neutron-star--black-hole systems, but is violated for comparable-mass black-hole systems of $\gtrsim 10 M_\odot$.

\subsection{Analytic approximation of the full likelihood}\label{sec:analytic-approx}

We can incorporate correlations of $f\sub{ISCO}$ with an arbitrary number of
waveform parameters by a suitable combination of the \FIM approximation
and the linearization introduced in (\ref{eq:logL_noiseless}). We follow the approach of Ohme {\textit{et al}}.\ \cite{Ohme:2013} who noted that in the case of a parameter-dependent cutoff frequency, the log likelihood (\ref{eq:logL_def}), with $d=h(\vtheta_0)$, can be expressed to leading order as 
\be
 \log L(\vtheta) \approx - \frac{1}{2} \left( \sum_{i,\,j} {\Gamma}_{ij} \Delta \theta_i \Delta \theta_j
+ 4 \left \vert \int_{f_1}^{f_2} \zeta(f) \,\dd f \right \vert \right).
\label{eq:logL_Fisher}
\ee
The first term constitutes
the standard (naive) \FIM approximation in the
frequency range up to the cutoff frequency of the reference signal, and the
second term accounts for the fact that the
(\ISCO) cutoff frequencies $f_1$ and $f_2$ 
can differ between the two waveforms.

The linear-order expansion of the second term is a generalization of
(\ref{eq:logL_noiseless}) and reads
\be
  4 \left | \int_{f_1}^{f_2} \zeta(f) \,\dd f  \right | \approx  4 \zeta(f\sub{ISCO}) \left|  \sum_i
 \frac{\partial f\sub{ISCO}}{\partial \theta_i}   \Delta
\theta_i  \right| . \label{eq:zeta_linear}
\ee
We recast the last expression in terms of 
\be
 \Delta \hat{\theta}_j \equiv 2 \zeta(f\sub{ISCO}) \sum_k
\left(\Gamma^{-1}\right)_{jk} \frac{\partial f\sub{ISCO}}{\partial
\theta_k},  
\label{eq:theta_hat}
\ee
leading to
\be
 4 \left | \int_{f_1}^{f_2} \zeta(f) \,\dd f  \right | \approx 2 \left | \sum_{i,\,j} 
\Gamma_{ij} 
\Delta \theta_i  \Delta \hat{\theta}_j \right |.  \label{eq:extranorm}
\ee
Combining (\ref{eq:extranorm}) with (\ref{eq:logL_Fisher}) and using the
symmetry of $\Gamma_{ij}$ finally leads to 
\ba
 \log L(\vtheta) &\approx&  - \frac{1}{2} \left( \sum_{i,\,j} 
\Gamma_{ij}  \Delta \theta_i \Delta \theta_j +  2 \left|
\sum_{i,\,j} \Gamma_{ij} \Delta \theta_i \Delta \hat{\theta}_j 
\right \vert \right) \\
   &=&  - \frac{1}{2} \sum_{i,\,j}  \Gamma_{ij} \left( \Delta
\theta_i \pm \Delta \hat{\theta}_i  \right) \left( \Delta \theta_j \pm \Delta
\hat{\theta}_j  \right) + \mathcal C.  \label{eq:logL_full}
\ea
The signs in the expression above are all positive when $\sum_{i,\,j} \Gamma_{ij}
\Delta \theta_i \Delta \hat{\theta}_j$  is positive, and negative otherwise.  
The constant $\mathcal C = \sum_{i,\,j}
\Gamma_{ij} \Delta \hat \theta_i \Delta \hat \theta_j/2$ simply
corrects the overall shift in $\log L$ introduced by completing the square.

\begin{figure}
\centering
 \includegraphics[width=0.73\textwidth]{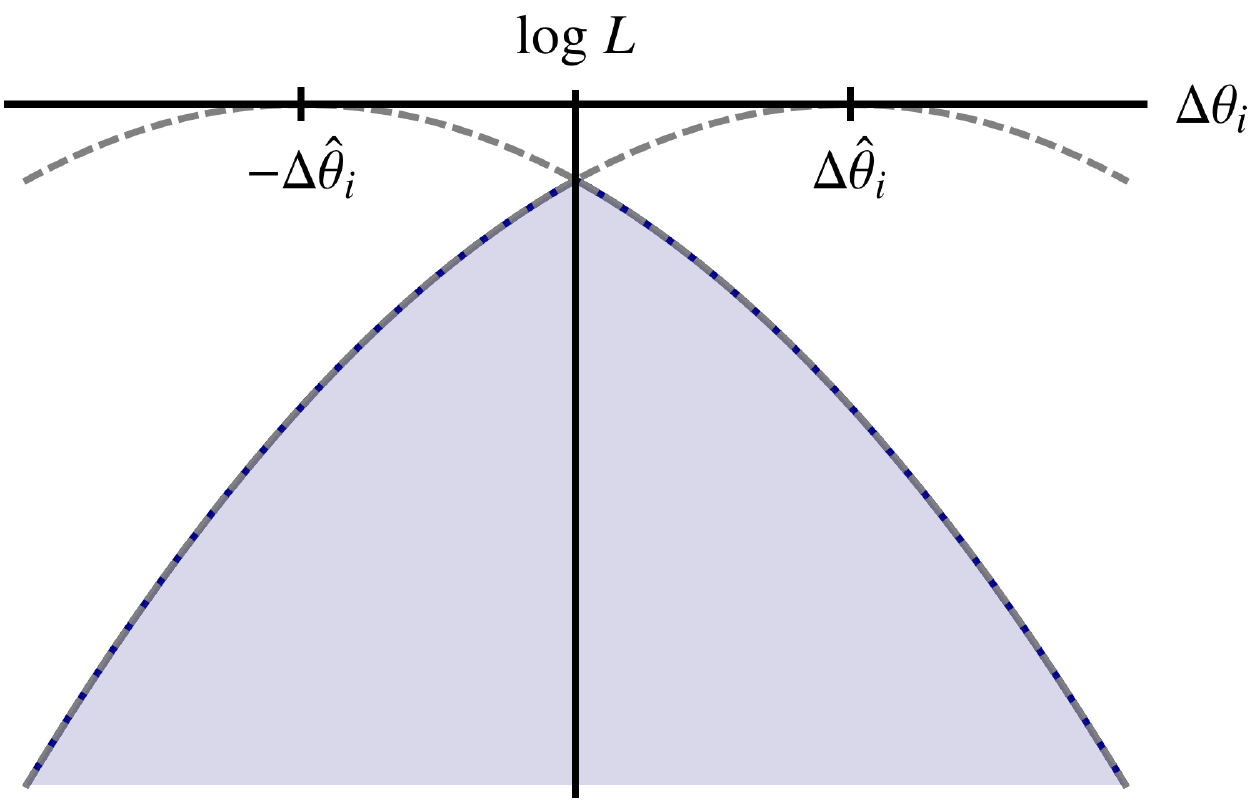}
 \caption{An illustration of the shape of $\log L$ as derived in
(\ref{eq:logL_full}), marginalised over all other parameters. An abrupt cutoff in the frequency domain manifests itself
at linear order as a symmetric displacement of the Gaussian likelihood by
$\Delta \hat \theta_i$, cf.~(\ref{eq:theta_hat}). The curve bounding the shaded
area constitutes the resulting log likelihood that is continuous, but not
differentiable at the origin.}
\label{fig:shape}
\end{figure}

Despite the issue of picking the correct sign for $\Delta \hat \theta_i$,
(\ref{eq:logL_full}) is easy to interpret. The marginalised likelihood over all but one parameter is a piecewise Gaussian function with the peak displaced by $\mp \Delta \hat{\theta}_i$, and
the sign is chosen such that the peak lies on the negative/positive axis for
positive/negative perturbations. A one-dimensional illustration is provided by
Fig.~\ref{fig:shape}. %

This combination of two displaced Gaussian functions has a sharp peak at the
maximum $\Delta \theta_i = 0$ and a faster fall-off around the peak than the
(naive) \FIM prediction.
Thus, parameter-estimation studies (such as \cite{Rodriguez:2013}) that properly
take the abrupt cutoff into account predict better parameter-estimation accuracy
than is achievable when such an artificial cutoff does not exist. Of course, if
the displacement $\Delta \hat \theta_i $ is small compared to the typical
parameter variations defined by the problem in question, then the overall effect
of an abrupt cutoff is small. In contrast, larger values of $\vert \Delta \hat
\theta_i \vert$ increase the effect of an abrupt parameter-dependent cutoff.

In our case, the natural parameter variation we should compare to is given by the standard deviation $\sigma_i$ of the (naive) \FIM prediction. The change in the likelihood function is significant only when the deviation $\Delta \hat \theta_i$ is significant relative to this variation. Hence, we can neglect the effect of a parameter-dependent cutoff frequency if
\be\label{eq:sigma_comparison}
 \vert \Delta \hat \theta_i \vert \ll \sigma_i,
\ee
which is a generalization of the toy case we discussed in the previous subsection.
For the simple case of a single parameter $M$, 
the condition (\ref{eq:ignore_cutoff}) becomes
\begin{equation}\label{eq:sigmaM_comparison}
|\Delta \hat M| \ll \sqrt{2} \sigma_M,
\end{equation}
consistent with (\ref{eq:sigma_comparison}).

One can draw several conclusions from these considerations and the explicit form of $\Delta \hat \theta_i$ given in (\ref{eq:theta_hat}). First, $\Delta \hat \theta_i$ is proportional to $\zeta(f\sub{ISCO})$. This confirms in full generality that an abrupt cutoff can be safely ignored if the waveform amplitude at this cutoff frequency is buried deep in the instrument noise.  This is the case for typical binary neutron-star systems observed with initial- and advanced-generation \GW detectors at reasonable \SNR.\footnote{The left-hand side of (\ref{eq:sigmaM_comparison}) is independent of distance with all other parameters fixed, while the right-hand side scales inversely with \SNR or linearly with distance, so the abrupt cutoff could still be significant for nearby low-mass sources. However, even for a binary with two $3 M_\odot$ components, this would only happen at $\rho \gtrsim 100$.} 

Furthermore, although the frequently-employed \ISCO cutoff frequency only depends on the total mass $M$ for binaries with non-spinning components, all parameters that are \emph{correlated} with $M$ are affected by the sharp cutoff as well. This becomes apparent in (\ref{eq:theta_hat}), where the gradient of the cutoff frequency is contracted with the inverse \FIM (which becomes the correlation matrix in the high-SNR limit). Conversely, parameters that are not significantly correlated with the total mass (e.g., the sky location in a network of identical detectors) are not affected. 

Last, both (\ref{eq:logL_full}) and Fig.~\ref{fig:shape} demonstrate that $\log L$ is \emph{not} differentiable at the origin, a fact that immediately puts the practicability and possible interpretation of a \FIM study into question. We discuss this in more detail in the next subsection. 

\subsection{Including an abrupt waveform cutoff in the \FIM}\label{sec:new_FIM}

In view of the above, it is perhaps not surprising that a full \FIM, which takes
all previously discussed effects into account, is not well defined
(at any \SNR) when an abrupt waveform cutoff is present. The \FIM, by
construction, predicts parameter uncertainties that scale inversely with $\rho$,
whereas information from an abrupt cutoff makes it possible to measure
parameters with an accuracy that scales inversely with $\rho^2$. Regardless of
the \SNR, it is impossible to accurately capture the character of the sharp peak
in the log likelihood with a quadratic function. The gradient of the log
likelihood is undefined at the true parameters when the log likelihood has a
sharp peak that renders its derivatives discontinuous. 

When the waveform has an abrupt cutoff, as in (\ref{eq:cutoff}), its derivative with respect to parameters includes an additional term proportional to the derivative of the Heaviside function, namely a delta function.  The full derivative $\tilde{h}_{,i}$ is 
\bel{partial}
\tilde{h}_{,i} = {\tilde{h}^0}_{,i} - \tilde{h}^0 \delta(f\sub{ISCO} - f) \frac{\partial f\sub{ISCO}}{\partial \theta_i},
\ee
where 
\be
{\tilde{h}^0}_{,i} = \left[ \frac{\partial A(f)}{\partial \theta_i}+ i A \frac{\partial \Psi(f)}{\partial \theta_i} \right] \exp\left[i \Psi(f)\right] H(f\sub{ISCO}-f) 
\ee
is the only part of the derivative included in standard calculations.

We can substitute (\ref{partial}) into (\ref{Gamma}) to calculate the full \FIM, which we denote $\check \Gamma_{ij}$. The full \FIM comprises the terms we have previously considered, the naive, inspiral-only $\Gamma_{ij} \equiv \inner{{\tilde{h}^0}_{,i}} {{\tilde{h}^0}_{,j} }$, and additional terms that occur whenever the cutoff $f\sub{ISCO}$ depends on the parameters $\theta_i$ or $\theta_j$. When the derivatives of $f\sub{ISCO}$ with respect to both $\theta_i$ and $\theta_j$ are non-zero, $\check \Gamma_{ij}$ includes a term that is proportional to the integral of a squared delta function, which is formally undefined, but can be considered infinite for this application. 

The naive \FIM $\Gamma_{ij}$, which is commonly used in calculations, is
incomplete for a waveform with abrupt cutoffs. This incompleteness can result
in apparent violations of the Cram{\'e}r--Rao bound. Rodriguez {\textit{et al}}.\ \cite{Rodriguez:2013} found
that the naive \FIM can overestimate uncertainties in mass parameters by a
factor of $5$--$10$ for binary black holes with total mass approaching $20
M_\odot$, i.e., in the regime where $\sigma\super{naive}_{f\sub{ISCO}} \gg
\sigma\super{abrupt}_{f\sub{ISCO}}$ (or, equivalently, $\Delta \hat \theta_i \gg
\sigma_i$ for parameter of interest $\theta_i$) and the information contents of
the cutoff is significant. On the other hand, the naive \FIM yields uncertainty
estimates compatible with a full Bayesian analysis that incorporates information from
the cutoff at binary neutron-star masses, where
$\sigma\super{naive}_{f\sub{ISCO}} \ll \sigma\super{abrupt}_{f\sub{ISCO}}$
($\Delta \hat \theta_i \ll \sigma_i$) and the information content of the
abrupt waveform termination is minimal (see Fig.~1 of \cite{Rodriguez:2013}).

Meanwhile, the full \FIM is ill-defined. Therefore, the \FIM cannot be used when
the waveforms have abrupt cutoffs. It is possible to regularize the \FIM by
replacing the step function with a more gradual taper, and progressively making
the taper more abrupt. As an example of this, we consider measurements of the
chirp mass 
\be
\mathcal{M} = \frac{(m_1 m_2)^{3/5}}{M^{1/5}}\,;
\ee
we find that the inverse \FIM
element $ (\check \Gamma^{-1})_{\mathcal{M} \mathcal{M}}$ can be reduced by many
orders of magnitude relative to the usual (and incomplete) calculation that only
considers $({\Gamma}^{-1})_{\mathcal{M}\mathcal{M}}$. For instance, for a binary
neutron-star system with $m_{1,\,2} =1.4 M_\odot$ and the waveform model and
noise model of \cite{Rodriguez:2013}, we find that the full $(\check
\Gamma^{-1})_{\mathcal{M}\mathcal{M}}$ is reduced by a factor of $\sim 10^{9}$
relative to the naive calculation. A complete \FIM with a regularized abrupt
cutoff does act as a Cram{\'e}r--Rao bound, resolving the apparent violation noted in
\cite{Rodriguez:2013}, but it is a useless extreme lower bound on the mass
parameters, underestimating the chirp mass uncertainty by a factor of $\sim
10^{4}$.

In the calculation above, we considered continuous integrals; in practice, data analysis relies on discretized data and templates. There are potentially two ways to discretize a waveform with an abrupt cutoff. In order to preserve the correct total waveform power, the waveform amplitude in the last non-zero frequency bin can be weighted to account for the value of the cutoff (\ISCO) frequency. This procedure yields the same results as in the continuous limit discussed above. Alternatively, if no weighting is used, the waveform is insensitive to changes of the cutoff frequency so long as the cutoff remains between frequency samples, and the likelihood is quadratic; a sample is abruptly added or removed from the waveform only when the cutoff frequency moves to a neighbouring bin, causing a discontinuous step in the likelihood. If the frequency-domain sampling step is larger than $\sigma\super{abrupt}_{f\sub{ISCO}}$, the naive \FIM calculation should apply; otherwise, the sharply peaked likelihood of Fig.~\ref{fig:shape} is replaced by a terraced shape, which yields the same limiting behaviour as the sampling step gets smaller.

\section{Abruptly terminating templates and complete signals}
\label{sec:fullsignal}

We detailed in the previous section how an abrupt signal termination in the
frequency domain affects the recovery of source parameters. The underlying
assumption that both signal and template contain such a cutoff yielded an
artificial increase in information. There are, however, only a few cases where
we would expect a real signal to terminate abruptly, e.g., coalescing
extreme-mass-ratio inspirals, where the plunge is rapid, and the merger and
ringdown have almost no power relative to the inspiral \cite{Amaro:2007}, 
or inspiralling stars that are tidally disrupted by their black-hole
companion before merger; and even in those cases the falloff could be abrupt in
the time domain, but not necessarily in the frequency domain. In most other
cases, we do not expect real \GW
signals to exhibit such an abrupt cutoff at all. Thus, the results derived
in Sec.~\ref{sec:cutoff} do not apply immediately to typical \GW searches.

Ideally, one should use waveform templates that actually model the
entire inspiral--merger--ringdown structure of the expected signals whenever a
sharp frequency cutoff has a considerable effect on the measurement as discussed
around (\ref{eq:sigma_comparison}). Nevertheless, there are cases where this is
either not practical or not possible, and performing \GW measurements
with accurate 
inspiral waveforms, neglecting merger and ringdown, can have considerable
computational advantages (e.g., \cite{ihope,S6PE}).
This section provides some discussion of how our previously derived
results change if abruptly terminating templates but complete signals are
considered. This allows us to determine an approximate region of mass space
for which the use of abruptly terminating waveforms for tasks such as template placement, gravitational-wave searches and parameter estimation is reasonable.

\begin{figure}
\centering
 \includegraphics[width=0.73\textwidth]{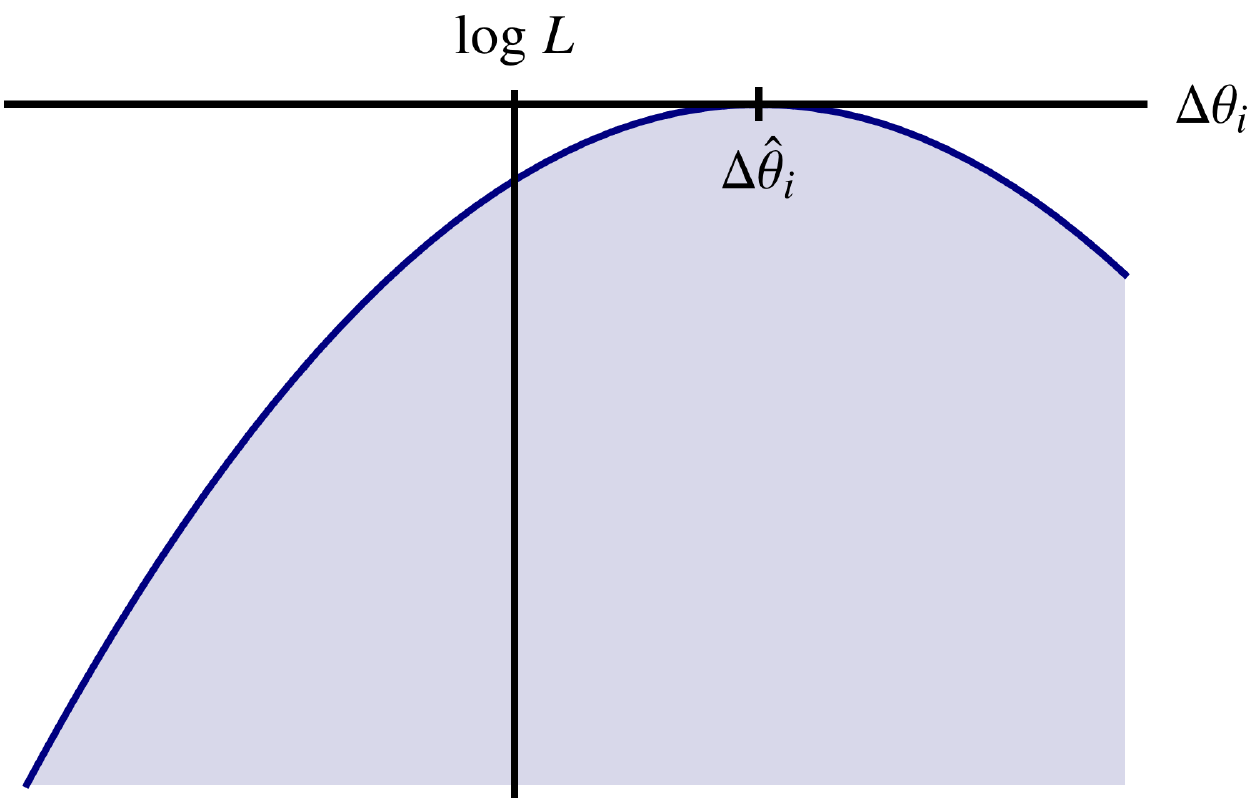}
 \caption{A one-dimensional illustration of the shape of $\log L$ as derived in
(\ref{eq:logL_full_noabs}). Parameter estimation on a complete signal with an abruptly terminated template family leads to a systematic bias $\Delta \hat \theta_i$, cf.~(\ref{eq:theta_hat}).}
\label{fig:shape2}
\end{figure}

To do so, we consider the best-case scenario, where the real signal and the
corresponding waveform template agree perfectly up to the template cutoff 
frequency $f\sub{ISCO}(\theta)$. We then find 
\be
 \log L(\vtheta) = -\frac{1}{2} \left( \inner{\Delta h}{\Delta h} \Big
\vert_0^{f\sub{ISCO}} + 4 \int_{f\sub{ISCO}}^\infty \zeta(f) \,\dd f \right),
\ee
which is similar to (\ref{eq:logL_Fisher}), although in this case the template
waveform is guaranteed to end at a lower frequency than the signal. An important difference, however, is that unlike the likelihood in (\ref{eq:logL_Fisher}) which contains an absolute value, this likelihood is everywhere smooth and differentiable (see Fig.~\ref{fig:shape2} and discussion below). Therefore, we expect to be able to carry out a Taylor expansion around the best-fit parameters and describe the covariance with an appropriate \FIM calculation.

The inner product $\inner{\Delta h}{\Delta h}$ is restricted to the frequency range where both
signal and template are nonvanishing, $f \in [0, f\sub{ISCO}]$.  We can
approximate it with the naive, inspiral-only \FIM that is well defined in this regime.
The second contribution has a nontrivial dependence on the actual shape of the
merger and ringdown amplitude, but we approximate it for small parameter
variations to first order as
\begin{eqnarray} 
 4 \int_{f\sub{ISCO}}^\infty \zeta(f) \dd f &\approx& \rho^2\sub{MR} - 4 \zeta
(f\sub{ISCO})  \Delta f\sub{ISCO} \label{eq:ISCO-template} \\
 &\approx& \rho^2\sub{MR} - 4 \zeta(f\sub{ISCO}) \sum_i \frac{\partial
f\sub{ISCO}}{\partial \theta_i} \Delta \theta_i ,
\nonumber
\end{eqnarray}
where $\Delta f\sub{ISCO} \equiv f\sub{ISCO}(\vtheta)-f\sub{ISCO}(\vtheta_0)$. 
The power contained in the merger and ringdown $\rho^2\sub{MR}$ quantifies the
loss in efficiency when we search with an incomplete (but
otherwise perfectly matching) model. It is constant in
this expansion and we neglect it as it is a pure scaling factor that
does not affect the shape of $L$.  

Then, by comparison with (\ref{eq:zeta_linear})--(\ref{eq:logL_full})
we find that
\be
 \log L(\vtheta) \sim - \frac{1}{2} \sum_{i,\,j} \Gamma_{ij} \left( \Delta \theta_i -
\Delta \hat \theta_i \right) \left( \Delta \theta_j -
\Delta \hat \theta_j \right) , \label{eq:logL_full_noabs}
\ee
where we have dropped all additional contributions that are
parameter-independent (again, they do not affect the shape of $L$) and $\Delta
\hat \theta_i$ is given by (\ref{eq:theta_hat}).

The obvious difference between (\ref{eq:logL_full}) and
(\ref{eq:logL_full_noabs}) is that in the case of a full signal,
the displacement vector $\Delta \hat \theta_i$ remains constant independent of
the sign of the parameter perturbations.
The reason is that previously, any deviation from the target cutoff frequency
resulted in an increase of the waveform difference. Now, however, an increase in
the template cutoff frequency can improve the match with the target
signal because a wider frequency range is covered by the template with
different parameters.

Fig.~\ref{fig:shape2} illustrates the behaviour of $\log L$ as derived in
(\ref{eq:logL_full_noabs}). The width of the likelihood or, more generally, the
parameter covariance, 
is the same as predicted by a naive \FIM that represents a smoothly terminating
inspiral-only calculation. We are generally justified in ignoring
second-order terms to the expansion
(\ref{eq:ISCO-template}), which formally constitute a correction to the naive
inspiral \FIM, as they are governed by small additions to the signal power from the
parameter-dependent cutoff, but we discuss the validity of this below. 
The main effect is a displacement of the peak of $\log L$ by $\Delta \hat \theta_i$, which can be identified as a
systematic bias caused by the disagreement between the full signal and the template waveform models. Since we have assumed that signal and template models match perfectly
up to $f\sub{ISCO}$, it is only the fact that the signal
extends to higher frequencies than the cutoff template which introduces the bias 
\be 
\Exn [\Delta \theta_i] = \Delta \hat \theta_i, \label{eq:sys_bias} 
\ee
where $\Delta \hat \theta_i$ are given by (\ref{eq:theta_hat}).

Let us consider the case where the template terminates at $f\sub{ISCO}$, which is a function of only the total mass as given by 
 (\ref{eq:fISCO}). Taking $\sigma_M^2 = (\Gamma^{-1})_{MM}$ we find
\be
 \Delta \hat M = - 2 f\sub{ISCO} \, \zeta(f\sub{ISCO}) \frac{\sigma^2_{M}}{M}. 
\label{eq:massbias}
\ee
where, as before $\zeta$ is the noise-weighted signal power as given in
(\ref{eq:zeta}).
As expected, a search with abruptly terminating templates tends to
underestimate the total mass to cover a larger frequency range. 

If we analyze a full signal with abruptly cutoff templates that faithfully reproduce the inspiral-only portion of the signal, we expect to find that both fractional statistical uncertainties and systematic biases on the mass parameter increase as the binary mass increases. The relative statistical uncertainty increases as fewer inspiral cycles are in the detector band for high-mass signals. The increase in the systematic bias is even more rapid because of the quadratic dependence on uncertainty in (\ref{eq:massbias}) and because the noise-weighted power $\zeta(f\sub{ISCO})$ increases as the cutoff frequency is lowered towards the most sensitive frequency region of the detector.

To illustrate the order of magnitude of this effect, we calculate inspiral-only
\acp{FIM} for binaries consisting of a $m_2 = 1.35 M_\odot$ neutron star and a black
hole with a mass $m_1$ between $5M_\odot$ and $20M_\odot$. All target systems have
non-spinning components, but our \FIM calculation takes variations of the
black-hole spin into account, although we constrain spins to be aligned with
the orbital angular momentum so that the system does not undergo precession. We
employ a frequency-domain post-Newtonian waveform model of the form
(\ref{eq:waveformmodel}), and details of our implementation are given in
\cite{Ohme:2013}. Although we are now considering the search for
complete signals, our first-order expansion framework does not require us to
use any specific merger--ringdown model. We assume a network \SNR of 10
(accumulated up to $f\sub{ISCO}$) for identical detectors characterized by the
Advanced LIGO noise curve in the zero-detuned high-power configuration
\cite{PSD:AL}, with a lower frequency cutoff at $10~\mathrm{Hz}$.

\begin{figure}
\centering
 \includegraphics[width=0.73\textwidth]{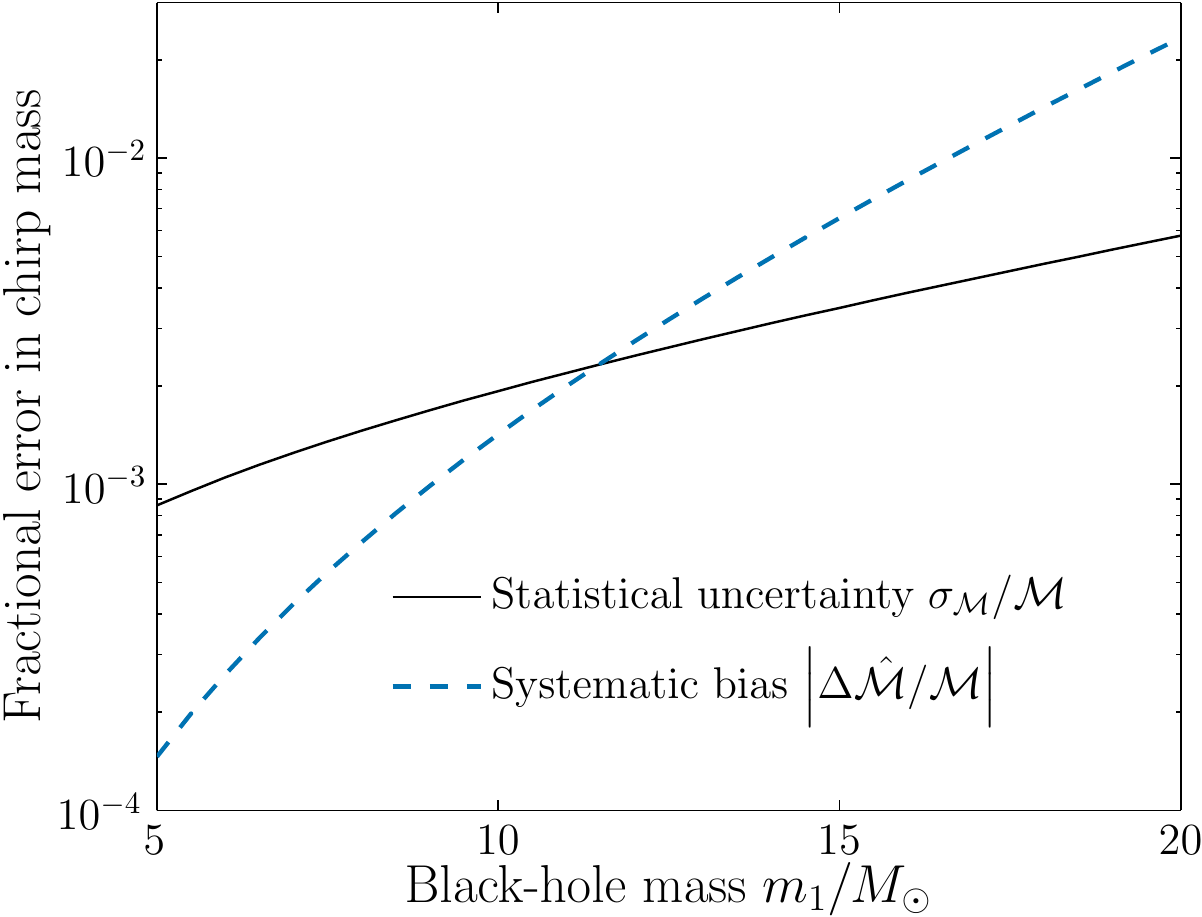}
 \caption{Relative statistical uncertainty $\sigma_\mathcal{M}/\mathcal{M}$
(solid line) and systematic bias $\Delta \hat{\mathcal{M}}/ \mathcal{M}$ (dashed
line) caused by using abruptly terminating templates, here shown for
non-spinning binaries with a $m_2 = 1.35 M_\odot$ neutron star and a black hole
with the indicated mass $m_1$ at $\rho = 10$. The underlying \FIM allows
variations of both masses, the black-hole spin (aligned with the orbital angular
momentum), as well as reference time and phase. The assumed detector is Advanced
LIGO (zero-detuned high-power configuration) with a lower cutoff at
$10~\mathrm{Hz}$ \cite{PSD:AL}.}
\label{fig:bias}
\end{figure}

In Fig.~\ref{fig:bias} we plot the statistical uncertainty in measuring the chirp mass
$\sigma_{\mathcal{M}}$ and the bias caused by an abruptly ending template
$\Delta \hat{\mathcal{M}}$. 
As expected, both the statistical uncertainty and systematic bias increase with increasing
mass. However, the systematic bias increases more rapidly and becomes the
dominant contribution at black-hole masses $m_1 \gtrsim 10 M_\odot$. 
Buonanno {\it et al.}\ \cite{PNwaveforms:2009} estimated the mass at which the merger and
ringdown contribute a significant fraction to the \SNR (specifically, where the effectualness drops to $0.97$) at $12 M_\odot$ for the same detector configuration. Our
calculation,
which is concerned with parameter bias for threshold signals, provides a similar
cutoff. For louder signals, the statistical uncertainties are reduced, such
that merger and ringdown can play an important part at lower masses.

Another related conclusion applies to template-placement algorithms that are
based on the \FIM normalized by the squared \SNR
\cite{Owen:1996,Keppel:2013,Brown:BNS,Harry:NSBH}. Commonly, such template banks
are constructed by requiring that signals and the nearest template have a match of at
least $0.97$. Our results show that the naive, inspiral-only \FIM prediction 
is still applicable for abruptly terminating templates, in case they are used
to search for complete signals. There is an additional systematic bias that, in
principle, does not harm the detection efficiency, but should be taken into
account when identifying the parameter range of the search or quoting the
parameters of the template with the highest data correlation. 

This systematic bias only becomes important when it
is comparable or larger than the template discretization, i.e., when 
$\Delta \hat \theta_i$ is outside the $0.97$ match region of the respective
template. For constant \SNR (which is a sufficiently good approximation of
optimized template \acp{SNR}) we
can use 
\begin{equation}
 \frac{1}{\rho^2} \inner{h(\vtheta_0)}{h(\vtheta)} \approx 1- \frac{1}{2\rho^2}
\sum_{i,\,j} \Gamma_{ij} \Delta \theta_i \Delta \theta_j \label{eq:match_fisher}
\end{equation}
(see \cite{Ohme:2013})
to relate a $0.97$ match on the left-hand side of (\ref{eq:match_fisher}) with a
statistical uncertainty predicted by the $\FIM$. We
find that the marginalized one-dimensional standard deviation $\sigma_i$
is equivalent to matches of $0.97$ if $\rho \approx 4$. Thus, if considering neutron-star--black-hole searches, one would have to increase the statistical uncertainty in Fig.~\ref{fig:bias}
accordingly to find that the systematic bias dominates in template banks for
black-hole masses $m_1 \gtrsim 17M_\odot$.

The above analysis was performed to leading order, and the example given for a specific binary source and detector configuration; thus, while it gives a reasonable rule of thumb as to when the systematic effect of neglecting merger and ringdown may become important, it should not be naively applied in all situations. In particular, any time the systematic error becomes comparable to the statistical uncertainty, expanding linearly around the true parameters is no longer guaranteed to produce meaningful results. Quadratic terms like $\dd\zeta / \dd f\sub{ISCO}\, (\Delta f\sub{ISCO})^2$ need to be included, though for sufficiently large biases, the expansion around the true parameters may not converge.

In summary, this calculation can explain systematic biases and predict their general trend (i.e., small masses are favoured), but the actual value sensitively depends on more
waveform features than those taken into account here. Moreover, while abrupt template termination does not introduce unwarranted measurement accuracy in the case when a full signal is considered and systematic biases are small, using incomplete templates without merger and ringdown contributions leads to an overly pessimistic prediction of measurement uncertainty.

\section{Conclusion 
\label{sec:conclusion}}

We have studied the influence of an abrupt waveform cutoff on parameter
estimation. Abrupt cutoffs are often used in \GW astronomy because of the
uncertainty in the merger and ringdown components of the waveform, or for ease
of computation. However, terminating the waveform can have undesired
consequences if this occurs in the band of the detectors, that is if there is
significant noise-weighted power at the cutoff frequency. It is therefore
desirable to use complete inspiral--merger--ringdown waveforms. If these are not
used, there are a number of effects to be aware of.

We have shown that there is potentially a significant amount of information encoded in the (in-band) abrupt termination of waveforms. They may appear to provide more information than is available in practice. Therefore, studies using abruptly terminated signals and templates may overstate the accuracy with which parameters can be recovered. In this paper, we evaluated the information contained in such a cutoff, determined when it was significant, and described how it could be approximately incorporated into an analytic calculation. Although this study was based on frequency-domain waveforms, we expect the abrupt termination of time-domain waveforms to yield analogous additional information in the cutoff; however, in practice, abruptly terminating time-domain waveforms are often tapered to avoid artifacts when transforming into the frequency domain, which ameliorates this effect.

The naive \FIM calculation is blind to the information encoded in the abrupt cutoffs.\footnote{All our analysis has been performed within a linearized framework. Therefore, many of our formulae are only directly applicable when considering small changes in parameters, just as the inverse \FIM can only be used to estimate the covariance in the linear-signal approximation.  Our colleagues \cite{ChoLee:2014} further developed the present work by demonstrating the impact of abruptly terminating templates within the effective Fisher Information Matrix formalism \cite{Cho:2013}.} This can create a difference between various approaches for measuring parameter-estimation accuracy when the same models are used, and can cause an apparent violation of the Cram{\'e}r--Rao bound. It also means that the naive, inspiral-only \FIM can give incorrect, overly pessimistic predictions if the physical model really does call for an abrupt cutoff.

While full parameter estimation with abruptly terminated waveforms incorporates unphysical information from the waveform termination, which can artificially improve parameter estimation accuracy, and the naive \FIM calculation avoids this problem, both ignore the information contained beyond the cutoff frequency, in the merger and ringdown phases of the waveform. Thus, there is a potential trade-off between the artificial gain of information from the sharp cutoff and the real loss of information from neglecting the merger and ringdown phases. 

The above results were obtained assuming that both the true signal and waveform
template include a cutoff. Using abruptly terminated waveform templates to
analyse a complete, non-terminating signal leads to a bias in the estimated
parameters. This is not surprising, since the templates do not match their
target signals. We have shown how to estimate the size of the bias, provided
that it is sufficiently small that the linearized framework remains valid.

In addition, we have shown that at lowest order the \FIM prediction of parameter covariances
remains unaffected by the template cutoff if the signal to be searched for
actually extends to higher frequencies. This proves, in hindsight, that previous
\FIM results in the literature are meaningful, even for heavier systems
containing black holes, if interpreted in the way outlined here. However, if the merger and ringdown power is significant, parameters could be extracted more accurately with full inspiral--merger--ringdown analyses than predicted by the naive, inspiral-only \FIM, so these studies may be overly pessimistic for massive systems.

An interesting application of this final result is to algorithms that build template banks for \GW searches based on inspiral-only \FIM predictions. Our analysis indicates that such banks still
cover the waveform manifold as desired if the underlying abruptly terminating
templates are used to search for complete signals. However, the loss in \SNR,
as well as the above-mentioned bias, generally remain unaccounted for. 

It is important to take care of the unintended consequences of using
unphysical models with sharp cutoffs. The best solution for obtaining accurate
estimates for parameter uncertainties lies in the use of waveforms that
faithfully capture the merger and ringdown phases, cf.\
\cite{PNwaveforms:2009,Ohme:2012,Sampson:2013,S6PE}.

\section*{Acknowledgments}
The authors are grateful to colleagues from the Birmingham \GW astrophysics
group, the Cardiff gravitational physics group and the LIGO--Virgo Science
Collaborations. This work was supported by the Science and Technology Facilities
Council and a Leverhulme Trust research project grant.

\section*{References}
\bibliographystyle{hunsrt.bst}
\bibliography{Fisher}

\end{document}